\documentclass[11pt]{article}
\usepackage{amsmath,amsfonts,slashed}
\usepackage{epsfig,epsf}
\usepackage{hyperref}

\def\be {\begin{equation}}
\def\ee {\end{equation}}
\def\beq {\begin{equation}}
\def\eeq {\end{equation}}
\def\bea {\begin{eqnarray}}
\def\eea {\end{eqnarray}}
\def\barr{\begin{array}}
\def\earr{\end{array}}

\def\opcit(#1){ {\em op. cit.}, #1}

\def\issue(#1,#2,#3){#1, #2 (#3)} 

\def\APP(#1,#2,#3){Acta Phys.\ Polon.\ \issue(#1,#2,#3)}
\def\ARNPS(#1,#2,#3){Ann.\ Rev.\ Nucl.\ Part.\ Sci.\ \issue(#1,#2,#3)}
\def\CPC(#1,#2,#3){Comp.\ Phys.\ Comm.\ \issue(#1,#2,#3)}
\def\CIP(#1,#2,#3){Comput.\ Phys.\ \issue(#1,#2,#3)}
\def\EPJC(#1,#2,#3){Eur.\ Phys.\ J.\ C\ \issue(#1,#2,#3)}
\def\EPJD(#1,#2,#3){Eur.\ Phys.\ J. Direct\ C\ \issue(#1,#2,#3)}
\def\IEEETNS(#1,#2,#3){IEEE Trans.\ Nucl.\ Sci.\ \issue(#1,#2,#3)}
\def\IJMP(#1,#2,#3){Int.\ J.\ Mod.\ Phys. \issue(#1,#2,#3)}
\def\JHEP(#1,#2,#3){J.\ High Energy Physics \issue(#1,#2,#3)}
\def\JPG(#1,#2,#3){J.\ Phys.\ G \issue(#1,#2,#3)}
\def\MPL(#1,#2,#3){Mod.\ Phys.\ Lett.\ \issue(#1,#2,#3)}
\def\NP(#1,#2,#3){Nucl.\ Phys.\ \issue(#1,#2,#3)}
\def\NIM(#1,#2,#3){Nucl.\ Instrum.\ Meth.\ \issue(#1,#2,#3)}
\def\PL(#1,#2,#3){Phys.\ Lett.\ \issue(#1,#2,#3)}
\def\PRD(#1,#2,#3){Phys.\ Rev.\ D \issue(#1,#2,#3)}
\def\PRL(#1,#2,#3){Phys.\ Rev.\ Lett.\ \issue(#1,#2,#3)}
\def\PREP(#1,#2,#3){Phys.\ Rept.\ \issue(#1,#2,#3)}
\def\SJNP(#1,#2,#3){Sov.\ J. Nucl.\ Phys.\ \issue(#1,#2,#3)}
\def\ZPC(#1,#2,#3){Zeit.\ Phys.\ C \issue(#1,#2,#3)}

\usepackage[dvips]{color}
\usepackage[normalem]{ulem}
\definecolor{darkgreen}{cmyk}{1,0,1,0.4}

\definecolor{darkred}{cmyk}{0,1,1,0.4}

\usepackage{slashed}

\newcommand{\lsp}{\tilde{\chi}^0_1}
\newcommand{\stopone}{\tilde{t}_1}
\newcommand{\stoptwo}{\tilde{t}_2}
\newcommand{\la}{\lambda}

\def \mlstop{m_{\stopone}}
\def \lspone{{\tilde\chi}_1^0}
\def \chonepm {{\tilde\chi}_1^\pm}
\def \met  {\mbox{${E\!\!\!\!/_T}$}}

\def\gsim{\buildrel{\scriptscriptstyle >}\over{\scriptscriptstyle\sim}}

\begin{document}

\renewcommand*{\thefootnote}{\fnsymbol{footnote}}
\begin{center}
 {\Large\bf{
LHC signatures of neutrino mass generation through R-parity violation}}

\vspace{5mm}

{\large Roshni Bose} \footnote{123.roshni@gmail.com}, 
{\large Amitava Datta} \footnote{adatta@iiserkol.ac.in},
{\large Anirban Kundu} \footnote{anirban.kundu.cu@gmail.com}\\[2mm]
{\em\small{Department of Physics, University of Calcutta, \\
92 Acharya Prafulla Chandra Road, Kolkata 700009, India}}\\[2mm]
{\large Sujoy Poddar} \footnote{sujoy.phy@gmail.com}\\[2mm]
{\em\small{Netaji Nagar Day College,
170/436 N.S.C. Bose Road, Kolkata - 700092, India.}}
\end{center}
\begin{abstract}

We consider a model of neutrino mass based on R-parity violating (RPV) 
supersymmetry, with three $\mu_i$, relevant for bilinear RPV terms, and 
three $\lambda'_{ijk}$, relevant for the trilinear terms. The present 
neutrino data, after a precise determination of the mixing angle $\theta_{13}$, 
severely constrain such models. We make a thorough study of one such 
class of models that may have interesting signatures at the upgraded LHC. 
In this class of models, the relevant trilinear couplings are of the form 
$\lambda'_{i33}$, so if the lighter stop squark $\tilde{t}_1$ is the 
lightest supersymmetric particle (LSP), it will decay only through these 
couplings, giving rise to events with isolated hard leptons and jets. Even when 
$\tilde{t}_1$ is the next-to-LSP (NLSP), it can decay via the tiny $\lambda'$ couplings
allowed by the neutrino data,  although it may face stiff competition from some 
R-parity conserving decay modes. Using a simple Pythia based simulation, we find 
that in both the cases the signal consisting of a pair of oppositely charged leptons 
($ee$, $\mu \mu$ or $e \mu $) plus jets may be observable at the upgraded LHC 
experiments for a reasonable range of the $\tilde{t}_1$ mass.
\end{abstract}

\date{\today}





\renewcommand*{\thefootnote}{\arabic{footnote}}
\setcounter{footnote}{0}

\section{Introduction}
\label{sec:introduction}

While the Standard Model (SM) has been vindicated \cite{sm-expt}, 
no less by the recent discovery of the 
Higgs boson \cite{atlas-higgs,cms-higgs}, as the correct theory of elementary particles 
at the electroweak scale, there are reasons to suspect that it is at most an effective 
theory, to be superseded by a more complete theory at a higher energy scale. One of the 
reasons is the tiny but nonzero neutrino mass, unaccounted for in the SM, whose existence 
has been inferred from the solar, atmospheric, and reactor neutrino experiments confirming the 
idea of neutrino oscillation \cite{pal}. The smallness of neutrino mass, and the apparent 
absence of right-handed neutrinos, associated with the fact that neutrinos, being charge-neutral, can 
very well be their own antiparticles, hints at the possibility that the neutrino mass terms
might be purely Majorana in nature. This in turn necessitates the presence of lepton number 
violating interactions, which are absent in the SM. 

Another reason to believe in the incompleteness of the SM is the fine-tuning problem of the 
Higgs boson mass \cite{ftuning}. Unlike the fermion and gauge boson mass terms, a scalar mass term does not 
spoil any symmetry of the action, so there is no reason why the quantum corrections to the 
mass would not drive it to the scale up to which the SM is valid. If the SM is valid all the
way up to the Planck scale $M_{Pl}=1/\sqrt{G_N} \sim {\cal O}(10^{19})$ GeV, 
this brings in a very unnatural 
fine-tuning in the theory, and this is considered a sufficient motivation for physics 
beyond the SM (BSM). While there are several options to avoid the fine-tuning problem, none 
of them are experimentally verified; however, supersymmetry (SUSY) 
\cite{susyrev} remains the most preferred option. 

Baryon number $B$ and lepton number $L$ are accidental 
symmetries of the SM and there is a priori no reason 
why the SUSY action would respect such symmetries. However, if both of them are violated, protons
would decay uncomfortably fast. To prevent that, one imposes an ad hoc $Z_2$ symmetry on the 
action, which is called the R-parity \cite{susyrev}, defined as $R_p = (-1)^{3B+L+2S}$ where $S$ is the spin 
of the particle, making $R_p=+1 (-1)$ for all particles (superparticles or sparticles). 
This forbids both $B$ and $L$ violating interactions as
well, and makes the lightest SUSY particle (LSP) stable, which can be a good cold dark matter 
(CDM) candidate \cite{DM}. This model is commonly known as the R-parity conserving (RPC) SUSY. 
The lightest neutralino ($\lspone$), a
weakly interacting massive particle (WIMP), is one of the most suitable CDM candidates.  

On the other hand, one might allow either $B$ or $L$ (but not both) violating 
interactions, which will still forbid the rapid proton decay but violate  R-parity. Such R-parity 
violating (RPV) models \cite{rpvrev,barbier} are obtained by augmenting the RPC SUSY Lagrangian by 
additional $L$ or $B$ violating terms. These terms lead to  
signatures that are drastically different from those of
RPC SUSY.  A typical example is the absence of large missing transverse energy (\met) signals, the 
hallmark of the RPC case, in the RPV models 
because of the unstable nature of the LSP (this, at the same time, means that in the RPV models 
even a colored or charged sparticle can be the LSP).

It turns out that $L$-violating RPV SUSY models can provide an excellent mechanism of generating 
Majorana masses for neutrinos, various facets of which have already been discussed in the literature 
\cite{Rakshit:2004rj,numass-rpvref,rpv-neutrino2}. However, the neutrino masses depend not only
on the RPV couplings but also on a number of parameters
in the RPC sector, like gaugino and squark or slepton masses, and the higgsino 
mass parameter $\mu$. The constraints on these 
parameters and the smallness of the neutrino mass
ensure that the required RPV couplings have to be quite small, typically $\sim {\cal O}(10^{-4})$ 
or smaller for sparticle masses of ${\cal O}(100)$ GeV \cite{abada}. This, in turn, means 
that RPV channels are going to be interesting only for the LSP decay; for all other cases 
(there is one exception that we will discuss below), they are going to be 
swamped by RPC decays, cascading down to the LSP. 

As we have just mentioned, the LSP can be a charged or colored sparticle in the RPV models. 
However, a well-motivated choice  is to have the lighter top squark (also called the stop) 
$\stopone$ as the LSP. This can happen because of the large top quark mass
which induces a significant mixing between the weak eigenstates of the stop,
which in turn tends to make one of the mass eigenstates lighter. This supports the possibility
of a $\stopone$ LSP. It may also be the next-to-lightest sparticle (NLSP),
with $\lsp$ being the LSP. In this paper, we will focus only on these two cases, namely, 
$\stopone$ LSP, and $\stopone$ NLSP.

If $\stopone$ be the 
LSP, it will decay through RPV channels with 100\% branching ratio (BR)
in spite of tiny RPV couplings as required by the models of neutrino mass. 
It is well-known that a class of RPV couplings that may generate the observed patterns for 
neutrino mass splittings and mixing angles
will result in  the signal $\stopone\to b+\ell$, where $\ell$ stands for
any one of the three charged leptons in the SM.
Thus, the signal for the stop pair production and decay will be an opposite sign dilepton
(OSDL) pair, accompanied by two hard jets and negligible $\met$ \cite{nmass,biswarup,naba,shibu}. 

It is worth emphasizing that it may be possible to observe the above RPV decays of the  $\stopone$ even if it is the 
NLSP while $\lsp$ is the LSP. Of course this can happen if some of the RPC decays of the 
$\stopone$ are either kinematically disallowed or dynamically suppressed. If the mass difference
between $\stopone$ and $\lsp$ is a little more than 100 GeV, such that the channel $\stopone\to bW \lsp$ opens up, 
the BRs 
for the RPV channels will go down drastically. Thus we restrict ourselves to the parameter space with
$m_{\stopone}-m_{\lsp} \leq 75$ GeV (this automatically rules out $\stopone\to t + \lsp$). 
In this scenario the competing RPC channnels like the flavor-changing 
neutral current (FCNC) decay \cite{hikasa} 
$\stopone\to c + \lsp$ and the four-body decay \cite{boehm} $\stopone\to f\bar{f} b \lsp$ are also suppressed. 
Unlike the stop LSP case, the combined BR of the RPV channels is not necessarily 100\%; it
depends on the parameters of the RPV and RPC sectors. 

The competition between the RPV and RPC stop decay modes were studied \cite{adspdsp} 
in the context of a model 
of neutrino mass characterized by three bilinear $\mu_i$ and three trilinear $\lambda'_{i33}$ 
RPV couplings ($i$ being the lepton generation index), 
 defined at the weak scale in a basis where the sneutrino 
 vacuum expectation values (VEV) are zero. 
It was found that the then neutrino data at the $2\sigma$ level 
could constrain the RPV parameter space pretty tightly; on the average 
only 4 per $10^6$ models, characterized by these 6 RPV couplings, passed the neutrino data. 
In Ref.\ \cite{adspdsp} the analysis was restricted to only a few benchmark scenarios for the RPC sector.

It was further shown that the neutrino oscillation data induces a typical hierarchy 
among the three $\la^{\prime}_{i33}$ couplings \cite{adspdsp}. Each  set of RPV parameters consistent 
with the oscillation data is characterized by one of the six possible hierarchies,
which is reflected in a similar hierarchy among the BRs of the RPV stop decays into 
the three leptonic channels. Thus a measurement of these BRs in  
colliders will provide a strong hint about  the underlying model of neutrino mass.  
The prospects of observing these decays at the Tevatron \cite{adsp1} and 
the LHC \cite{adsp2} were estimated by two of us using {\tt PYTHIA} \cite{pythia} 
based analyses.
 
In view of the fact that the experiments at LHC operating at $\sqrt{s}=14$ 
TeV are round the corner, we update 
and upgrade the analysis of \cite{adsp2} in the following ways. 
We use the latest neutrino data \cite{rpv-nu-fit} which have undergone a considerable 
change compared to the one used in Ref.\ \cite{adsp2}. A striking example is $\theta_{13}$, now
definitely known to be nonzero, for which only an upper bound was used in Ref.\ \cite{adsp2}. 
At $1\sigma$ level, $\sin^2\theta_{23}$ is now split into two allowed regions excluding the 
erstwhile canonical value of $0.5$, and the spread is now significantly above or below $0.5$. 
A fresh look at  the allowed parameter space (APS) consisting of all combinations of the six RPV
parameters consistent with the neutrino data \footnote{Each combination in this set will be refered
to as a solution.} 
and the corresponding LHC signals is, therefore, called for.
We also systematically study the effect of variation of the RPC parameters on the 
APS and document the results, while only a few benchmark RPC points were 
considered in the earlier analysis, on which the RPV 
parameters were scanned to find the APS.  Last but not the least, now that the Higgs boson mass has been measured, 
a large stop mixing parameter $X_t$ is mandatory unless one is ready to introduce 
superheavy squarks. This parameter augments the mixing between the stop weak eigenstates, 
making the light $\stopone$ scenario even more appealing.

In the LHC related analysis, we consider much higher values of $\mlstop$ 
compared to Ref.\ \cite{adsp2} for obtaining a rough estimate 
of the mass reach. Since neither $b$- nor $\tau$-tagging efficiency at LHC-14 TeV is well documented
as yet, we do not include them in our analysis. 
For signals involving $\tau$ there is an additional uncertainty; a serious background may arise 
due to pure QCD processes where jets can be mistagged as $\tau$-jets. It is therefore better to wait 
for a data driven estimate of this background as and when it is available. 
We have thus restricted ourselves to final states only with $e$ and/or $\mu$, and 
do not consider the $b$-tagging.
As we shall see, even with this simple minded approach (which can be further 
improved as and when the selection procedures and data driven
background estimates from the LHC collaborations are available), a lot of parameter space can be covered.

Signals for stop decays mediated by the single coupling $\lambda'_{333}$ 
($\stopone\to b+\tau$) have been explored 
by the CMS collaboration \cite{rpv-cms-2012}. The bound does not 
depend on the exact values of the RPC or RPV couplings if
$\stopone$ is the LSP; the BR is always 100\%. 
In this case the CMS group found a lower limit of  453 GeV on the mass of $\stopone$.
In a model of $\nu$ mass leading to $\stopone$ decays, one requires three nonzero 
$\lambda'_{i33}$ couplings; thus, the CMS bound
is valid only in the special case $\lambda'_{333} \gg \lambda'_{133}, \lambda'_{233}$. 
For other hierarchies among the $\lambda'_{i33}$s the above bound will get diluted. 
We have 
restricted ourselves, as just mentioned, to $\stopone$ decays into final states with $e$ or $\mu$.
Since the detection efficiencies of $e$ and $\mu$ are likely to be greater than 
that of $\tau$, we work with the conservative bound $\mlstop \gsim$ 500 GeV.

The paper is arranged as follows. In Section II, we briefly summarize the neutrino mass generation 
mechanisms through RPV interactions. The next section deals with the 
signals of $\stopone$ decays in the RPV channels mentioned above and 
the corresponding SM backgrounds following Ref.\ \cite{adsp2}. 
We also estimate the minimum observable product branching ratio (MOPBR), 
to be defined in Section III, for these channels as a function of $\mlstop$. 
In Section IV we discuss our analysis, based on the random generation of $10^{9}$ RPV models 
for both stop LSP and stop NLSP cases, and filtering them through the latest neutrino 
data. We compute the product branching ratios (PBR) for the relevant 
RPV decays of $\stopone$ in the models consistent with the data and check the observability 
of the resulting signals using the MOPBRs derived in Section III.  We also study 
the impact of variation of the important RPC parameters related to 
the neutrino mass matrix on the allowed solutions. In Section V, we summarize and conclude.

\section{Neutrino mass with RPV}

We consider a model with three bilinear R-parity violating (RPV) couplings $\mu_i$ 
and three trilinear couplings $\lambda'_{i33}$ at the weak scale, so that the RPV 
part of the superpotential looks like
\be
{\cal W}_{\rm RPV} = \mu_i L_i H_2 + \lambda'_{i33} L_i Q_3 D^c_3\,,
\ee
where $L_i$, $Q_3$, $D_3$ and $H_2$ are the lepton, third-generation SU(2) doublet 
quark, SU(2) singlet bottom quark, and the second Higgs doublet (the one responsible 
for giving mass to the top quark) superfields 
respectively, and $i=1,2,3$ is the generation index. 
All trilinear RPV couplings are taken to be real; making them complex does not affect 
the result. 

In general, two bases for the 
$\{H_2,L_i\}$ superfields  are used in the literature \cite{barbier}. One is where 
all the neutral components, the Higgs and the sneutrinos alike, get nonzero VEV. 
One can also rotate the basis in such a way that the sneutrino VEVs are zero in the 
new $\{H'_2,L'_i\}$ basis, but the fields here
are combinations of fields of the first basis. We choose the basis where the sneutrino 
VEVs are zero. It turns out that the rotation, 
which is a measure of the misalignment (or the sneutrino component of the Higgs),
is extremely tiny, governed by $\mu_i/\mu$. To satisfy the neutrino data, one needs
$\mu_i\sim 10^{-4}$ GeV, and if we take $\mu\sim 1$ TeV, $\mu_i/\mu \sim 10^{-7}$
(see the end of section IV for a more quantitative estimate).
As a result, we have retained this mixing where it is the dominant effect ({\em e.g.} 
neutralino-neutrino mixing which directly contributes to the neutrino mass matrix) but have neglected this 
where it induces a subdominant effect ({\em e.g.} see the discussion on stop decay in Section IV).


The most general neutrino mass matrix can be found, {\em e.g.} in \cite{barbier}. 
Due to our choice of basis one may drop the terms containing the sneutrino VEVs. 
Possible additional contributions due to mixing between different fields are strongly suppressed due to mixing effects 
discussed in the last paragraph.

In principle the bilinear soft RPV terms $B_i$ may contribute to
the neutrino mass matrix. However, in view of the fact that there are too many free parameters in RPV SUSY 
models, it is a standard practice to work with a minimal set of RPV parameters which can adequately describe 
a physical phenomenon (neutrino mixing in the present case). In spirit it is like employing a `simplified model' 
in the context of RPV SUSY used by the LHC collaborations and also by a number of  phenomenological analyses. 
Following this practice we have assumed the $B_i$'s to be negligibly small.

Using the above approximations,  
the neutrino mass matrix $M_{\nu}$ has contributions coming from both tree and loop level processes as listed below.
Bilinear RPV induces a mixing between neutrinos and neutralinos, and the 
tree-level neutrino mass matrix is
\be
M^{\rm {tree}}_{ij}=\frac {M_Z^2 M_{\tilde\gamma} \cos^2\beta} 
{\mu\left(M_Z^2 M_{\tilde\gamma}\sin 2\beta - \mu M_1 M_2\right)} \, \mu_i \mu_j 
                  \equiv C  \mu_i\mu_j\,.
\label{mtree-c}
\ee
Here $M_1$ and $M_2$ are the $U(1)$ and $SU(2)$ gaugino masses, $\mu$ is the 
higgsino mass parameter, 
$M_{\tilde\gamma} = M_1 \cos^2\theta_W + M_2 \sin^2\theta_W$, and 
$\tan\beta = v_2/v_1$ is the usual ratio of the two Higgs VEVs.
The flavor indices are denoted by $i,j$.
The first term in the denominator of Eq.\ (\ref{mtree-c})
is much smaller compared to the second if the lightest neutralinos are gaugino 
dominated ($\mu \gg M_1, M_2$), or $\tan\beta \gg 1$. In either of these limits, we can write
\be
M^{\rm {tree}}_{ij}= - \frac {M_Z^2 M_{\tilde\gamma} \cos^2\beta} 
{M_1 M_2} \, \frac{\mu_i}{\mu} \, \frac{\mu_j}{\mu}  
                  \equiv -C' \, \frac{\mu_i}{\mu} \, \frac{\mu_j}{\mu}\,.
                  \label{mtree-cpr}
\ee
This approximate formula is valid throughout this paper, governed by our choices of the 
SUSY parameters. Note that $C$ is negative if the 
denominator of Eq.\ (\ref{mtree-c}) is negative, and this statement is 
independent of the sign of $\mu$ in the limit for which Eq.\ (\ref{mtree-cpr})
is valid. 

This mass matrix has obviously only one non-zero eigenvalue, and is therefore unable
to explain the neutrino data, which needs at least two non-zero mass eigenvalues.
With the introduction of the trilinear RPV couplings, 
a variety of other diagrams contribute to the neutralino-neutrino mass matrix
at the one-loop level \cite{Rakshit:2004rj,numass-rpvref},
and the new entries to the neutrino mass matrix look like 
\bea
M^{\lambda'\lambda'}_{ij}&=&
\frac{3}{8\pi^2}\sum_{k,l} \frac{A_d-\mu\tan\beta} { m^2_{\tilde d_k}} \lambda'_{ilk}
                \lambda'_{jkl} m_{d_l} m_{d_k}\,,\nonumber\\ 
M^{\mu\lambda'}_{ij}&=&
\sum_k\frac{3}{16\pi^2}g_2 m_{d_k}\frac{\mu_i\lambda'_{jkk}}{m_{\tilde d_k}}\,.
\label{rpv-numass-loop}
\eea
Here $m_{\tilde d_k}$ is the average down-type squark mass parameter, $A_d$ 
is the relevant trilinear coupling,  and $m_{d_i}$ is the 
$i$-th generation down-type quark mass. Note that Eq.\ (\ref{rpv-numass-loop}) 
is a bit simplified expression, {\em e.g.} we neglect the quark flavor mixing and approximate 
the squared mass difference of the two squark mass eigenstates by the product of their average 
mass and the corresponding quark mass. For simplicity, we neglect the soft R-parity breaking 
bilinear $B_i$ terms and the corresponding entries to the neutrino mass matrix generated 
by them. 

We will focus only on the cases where the lightest stop, $\stopone$, is the LSP (or NLSP),
and decays exclusively (mostly) through RPV channels, yielding a lepton in the final state. 
Thus, one of the indices $k$ and $l$ in Eq.\ (\ref{rpv-numass-loop}) must be 3; leading to
three possible combinations: $\lambda'_{i31}\lambda'_{j13}$, $\lambda'_{i32}
\lambda'_{j23}$, and $\lambda'_{i33}\lambda'_{j33}$. We work in a framework where the 
first two generation squarks are taken to be much heavier than the third generation ones
and therefore the matrix element has a suppression from the squark propagator if
the combination $\lambda'_{i31}\lambda'_{j13}$ or $\lambda'_{i32}
\lambda'_{j23}$ is chosen. Moreover, for these two combinations
the quark mass factors, coming from the mass insertion in the internal quark line and the 
mixing in the internal squark line, will also be smaller ($m_dm_b$ or $m_sm_b$) 
compared to the choice $\lambda'_{i33}\lambda'_{j33}$, where the corresponding factor is $m_b^2$.
As a result much larger values of the RPV couplings compared to the $\lambda'_{i33}
\lambda'_{j33}$ case would be required to generate the same   $M^{\lambda'\lambda'}_{ij}$. 
However, the combinations like $\lambda'_{i1(2)3}\lambda'_{i31(2)}$ contributing
to the diagonal elements of the mass matrix are severely restricted from 
tree-level $B_d$--$\bar{B_d}$ ($B_s$--$\bar{B_s}$) mixing \cite{ak-jps}, the product being 
of the order of $10^{-6}$ or smaller for 300 GeV sleptons. This tension leads us 
to the last set, introducing three RPV 
couplings $\lambda'_{133}$, $\lambda'_{233}$ and $\lambda'_{333}$ in the model
 \footnote{We will be interested in the $b$-quark 
only, for which we take the mass to be $m_b(m_b) = 4.25$ GeV.}.

This removes the summation over the quark generation 
indices in Eq.\ (\ref{rpv-numass-loop}), and both $M^{\lambda'\lambda'}$ and $M^{\mu\lambda'}$,
now given by
\bea
M^{\lambda'\lambda'}_{ij}&=&
\frac{3}{8\pi^2}\frac{A_b-\mu\tan\beta} { m^2_{\tilde b}} m_b^2 \lambda'_{i33}
                \lambda'_{j33} \equiv K \lambda'_{i33}\lambda'_{j33}\,,\nonumber\\ 
M^{\mu\lambda'}_{ij}&=&
\frac{3}{16\pi^2}g_2 m_b\frac{\mu_i\lambda'_{j33}}{m_{\tilde b}}\,,
\label{rpv-numass-loop2}
\eea
become rank-1 matrices with only one non-zero eigenvalue. However, when we add all three 
terms, there are two nonzero eigenvalues (as all the matrices are not diagonalizable by the 
same unitary transformation), the first one is at about the same place as that 
of the tree-level $C\mu_i\mu_j$ matrix, and another one which is much smaller than that. 
Thus, the separation between them gives the atmospheric splitting, and the separation between 
the small nonzero eigenvalue and zero gives the solar splitting, leading to a normal 
hierarchy. 

That the second eigenvalue 
is small can be ascribed to the choice of the parameters that one employs. 
However, there is a way to see that the normal hierarchy is more natural in this scenario. Suppose the
RPC and RPV parameters are such that $M^{\lambda'\lambda'}$ gives a large eigenvalue, of the 
order of $0.1$ eV, and the contributions from the other two matrices, $M^{\rm tree}$ and 
$M^{\mu\lambda'}$ are small. This will still give a normal hierarchy. Only in the case of a
nontrivial fine-tuning among all these parameters we expect an inverted hierarchy, two large 
eigenvalues that are almost degenerate. Also, this has to pass the constraints 
on the neutrino mixing angles which is already quite precise. In fact, we did not 
get any solution that supports inverted hierarchy even if we play with all the SUSY parameters.

\section{The signals and the SM backgrounds}

In this section we study the prospect of detecting the RPV decays 
of $\stopone$ at the upgraded LHC by a simple minded simulation following Ref.\ \cite{adsp2}. 
Our analysis, however, is extended to  much larger values of $\mlstop$.

As already discussed we restrict ourselves to  the RPV decays of 
$\stopone$ through the couplings $\lambda'_{133}$ and $\lambda'_{233}$ 
only, focussing on electrons and muons in the final state and neglecting the $\tau$ leptons. 
The signal size is sensitive to $P_{ij}$,
the product branching ratio (PBR) for the $ij$-th channel,  given  by
\be
P_{ij} = {\rm Br}\, (\stopone \to \ell_i + b) \times {\rm Br}\, 
(\stopone \to \ell_j + b)
\label{Pij-def}
\ee
where $i,j=1(2)$ for $e(\mu)$.
Note that we are considering the decay of the mass eigenstate $\stopone$, 
whereas the weak eigenstate $\tilde{t}_L$ takes part in RPV decays. Thus, the 
amplitudes of $\stopone$ decay depend on the stop mixing angle $\theta_t$, defined as
\be
\stopone = \tilde{t}_L\cos\theta_t + \tilde{t}_R\sin\theta_t\,,\ \ 
\stoptwo = -\tilde{t}_L\sin\theta_t + \tilde{t}_R\cos\theta_t\,.
\label{theta-t-def}
\ee
However, if $\stopone$ is the LSP, there is no RPC decay channel open for it, so the $\theta_t$ 
dependence cancels out in the BRs and hence in the PBR. This is not so
for the $\stopone$ NLSP case.

In this section we treat the $P_{ij}$s as model independent free parameters and   
estimate the minimum value of $P_{ij}$ corresponding to observable signals in  $ee$, $\mu \mu$ and $e \mu$  
channels. In the next section we compute $P_{ij}$ in different randomly generated models of 
neutrino mass, 
and see how many of them would lead to observable signals in different channels.

We simulate the pair production of the lightest stop, $\stopone$, at 14 TeV LHC 
using {\tt PYTHIA} \cite{pythia}. For this analysis, initial and final state radiation, 
decay, hadronization, fragmentation and jet formation are implemented according to the 
standard procedures in {\tt PYTHIA}.

For reconstruction of jets we follow the cone algorithm procedure described in
Ref.\ \cite{adsp2} using {\tt PYCELL} in {\tt PYTHIA}.
Final state $e^\pm$ and $\mu^\pm$ are selected with transverse 
momentum $P_T \geq 20$ GeV and $\vert\eta \vert < 2.5$.
To suppress the SM backgrounds the following selection criteria (SC) are used:

\begin{itemize}
\item  SC1: For isolated $e$ and $\mu$ we require lepton-jet angular separation,
 $\delta(l,j) >0.5$. The $e$ and $\mu$ identification efficiencies are 
assumed to be 100\%.

\item SC2: Events with two isolated leptons ($e$ and $\mu$) with $P_T >150$ GeV are 
selected. 
\item SC3: Only events with two jets are selected. 

\item SC4: Two highest jets are required to have $P_T>100$ GeV.

\item SC5: Only events with missing transverse energy $E_T\!\!\!\!\!\!\!\slash \,\,\,\,< 60$ GeV are kept.

\end{itemize}
We reiterate that in this analysis $b$-tagging is not implemented.

We consider backgrounds arising from $t \bar t, WW, WZ, ZZ$,
Drell-Yan (DY) and QCD events. The stronger $P_T$-cut on leptons (SC2) 
dramatically reduces all backgrounds; except $t\bar t$ all other backgrounds 
become vanishingly small. Moreover, for $ t \bar t, WW$ and $WZ$ events SC5 
severely suppresses all of them.  
For further details about the cuts, we refer the reader to Tables 4, 5 and 6 of Ref.\  
\cite{adsp2}. In Table 4 of \cite{adsp2}, it was shown that $t \bar t$ events 
significantly reduce to a small value after SC5 although the maximum suppression 
comes from  SC2.  

Among the relevant backgrounds QCD is very large, because 
the QCD cross-section is very large. The leading order cross-sections 
were computed by {\tt PYTHIA} in two $P_T$ bins:
(i) 400 GeV $< P_T <$ 1000 GeV and (ii) 1 TeV $< P_T <$ 2 TeV. 
The QCD scale has been set to be $\sqrt s$. 
The cross-sections corresponding to the above two bins are 2090 pb and 
10 pb respectively. Beyond 2000 GeV, the number of events are negligible, and so
we do not consider them here. Due to SC2 QCD events drop down to zero (see 
Table 5 of \cite{adsp2}).

The cut SC2 brings the DY events down to a very low level, and SC3 and SC4 finally make it 
negligible. Similarly, SC2 makes $WW$, $WZ$, and $ZZ$ events vanishingly small. SC5 plays 
the same role for $t\bar{t}$ events; it is also effective for $WW$ and $WZ$.

In Table 6 of \cite{adsp2} all SM backgrounds are presented for the integrated luminosity 
${\cal L} = 10$ fb$^{-1}$.
However, in this work we consider ${\cal L}$ to be either 100 fb$^{-1}$ or 300 fb$^{-1}$,
so the numbers given in that table are to be multiplied by 10 or 30 respectively. 
While the few surviving events were
eliminated with $b$-tagging in \cite{adsp2}, we do not use this for our analysis as the signal is already sharp 
enough.

For a given integrated luminosity ${\cal L}$ the minimum observable product branching ratio
(MOPBR) for each process is given by
\begin{eqnarray}
P_{ij}^{\rm {min}}& = &\frac{5 \sqrt {\eta\,{\cal L}\, \sum_y \left(\sigma_y \epsilon_y\right) }}
{\eta \, {\cal L}\, \sigma(\stopone \stopone^\ast) \epsilon}\,,
\label{def-pbr}
\end{eqnarray}
where $P_{ij}$ is defined in Eq.\ (\ref{Pij-def}),
$\sigma_y$ and  $\epsilon_y$ denote, respectively, the cross-section and the efficiency 
of the background of type $y$.
The final efficiency for the signal is denoted by $\epsilon$.
The factor $\eta$ is equal to $1(2)$ for $i = j$ $(i \not= j)$. In essence, this corresponds 
to a $5\sigma$ significance for discovery. 

\begin{figure}[!htbp]
\begin{center}
{
\epsfxsize=6cm\epsfbox{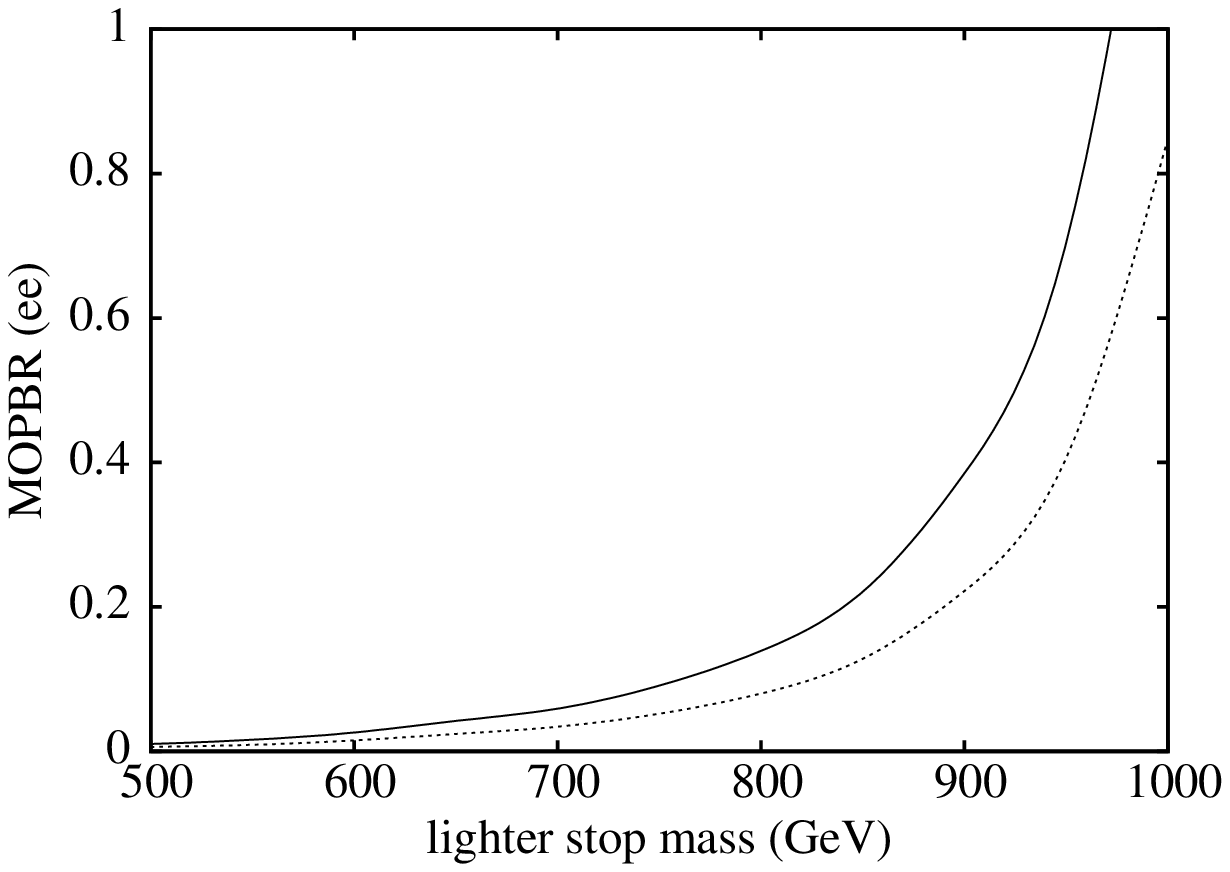}
}
\hspace{-8mm}
{
\epsfxsize=6cm\epsfbox{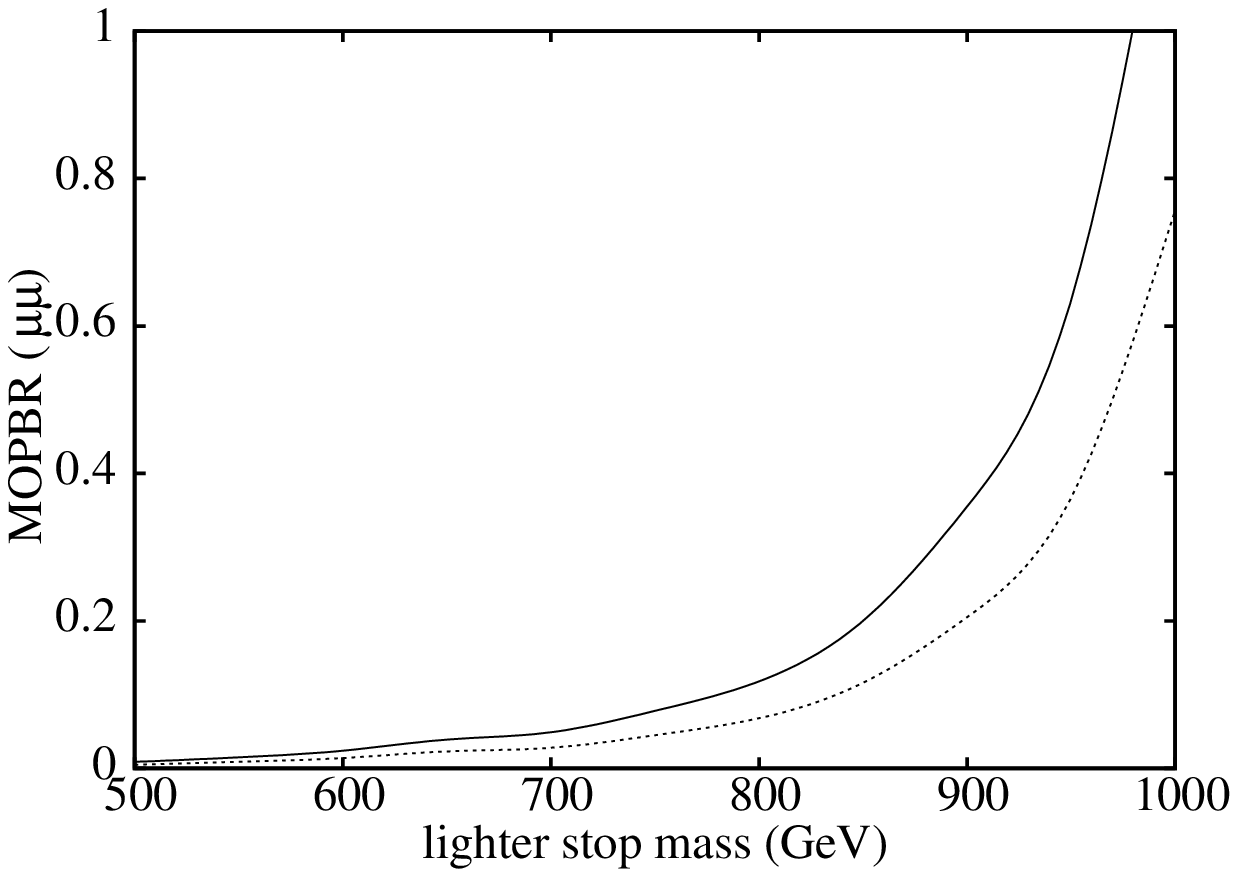}
}
\hspace{-8mm}
{
\epsfxsize=6cm\epsfbox{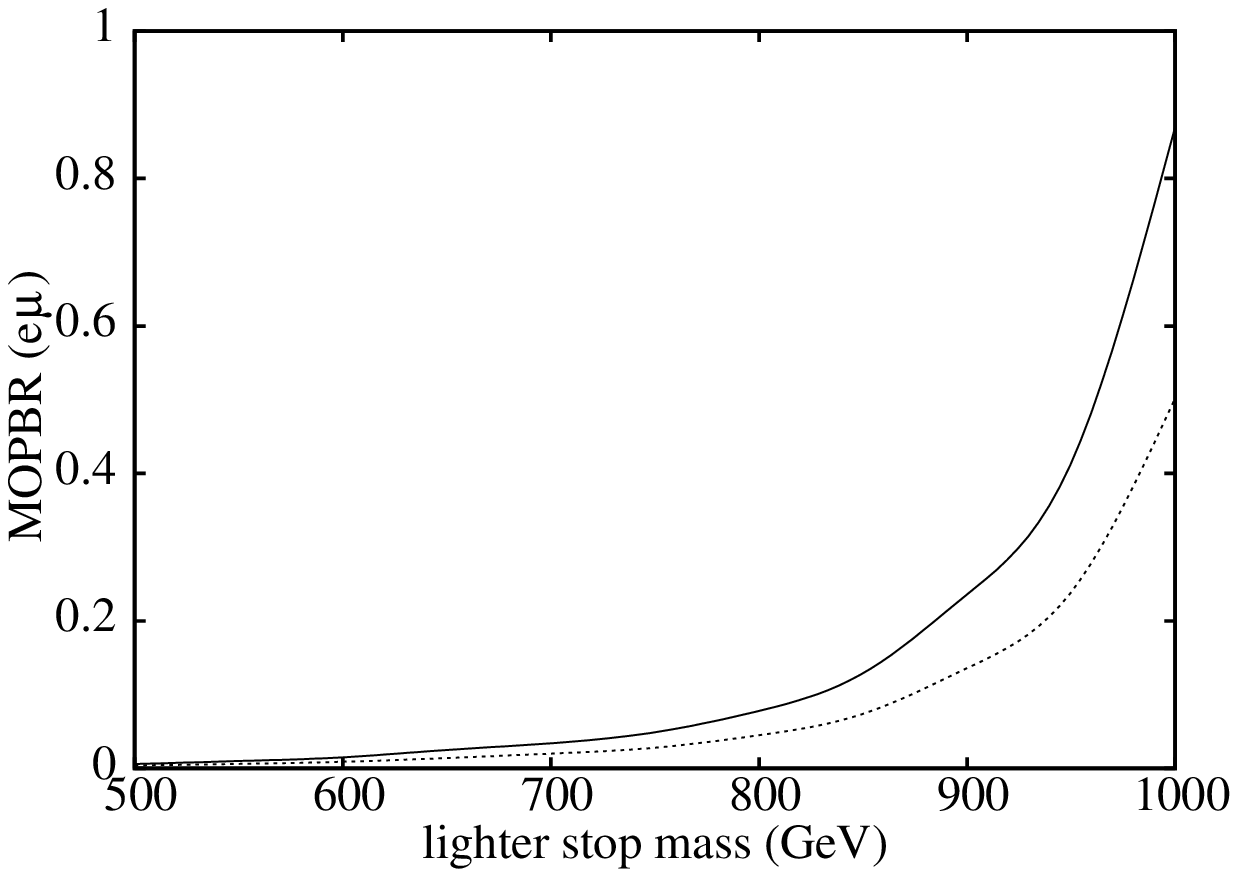}
}
\end{center}
\caption{The MOPBR lines for the channels $e^+e^-$, $\mu^+\mu^-$, and $e^\pm \mu^\mp$ 
coming from the RPV decays of the lighter stop pair. The upper red (lower blue) curve is for 
an integrated luminosity of 100 (300) fb$^{-1}$. A signal can be observed if the PBR falls 
above the MOPBR line.}
    \label{fig:lsp-reach}
\end{figure}
In Figure \ref{fig:lsp-reach} we plot the model independent MOPBR as a function of $\mlstop$
for different RPV decay channels of $\stopone$. The upper (lower) curve corresponds to
an integrated luminosity 
${\cal L} =$ 100 (300) fb$^{-1}$. For any solution, a particular signal will be observable if
the PBR in that channel exceeds the correponding MOPBR.  
Note that the higher luminosty option plays a marginal role for low $\mlstop$ ($< 600$ GeV), 
and again has very little to do if the kinematic threshold is reached for high $\mlstop$ 
($> 950$ GeV). The improvement can be seen only in the intermediate region.

One might wonder whether, for the stop NLSP case, the RPV decays of the LSP $\lsp$ can fake the signal 
($q \bar{q} l \bar{l}$ without large $\met$). However, these decays will be a serious background only 
if the $\lsp$ is directly pair produced. If it is produced in association with other sparticles or 
come from a decay cascade of heavier sparticles, the multiplicity of leptons and jets in the final state 
will be very different from our signal. Thus a significant background can arise only with LSP pair 
production followed by the RPV decays of $\lsp$ into $l + W$ and  $\nu + Z$. It may be noted that 
the LSP pair production cross-section at the LHC is negligible compared to the stop pair production. 
For example, with $m_{\lsp}=525$ GeV and $m_{\stopone}=600$ GeV, the cross-sections of stop pair 
production and LSP pair production are $0.44$ pb and $0.27$ fb respectively. Since the BRs 
involved in the signal and backgrounds are $\sim {\cal O}(1)$, the fake events from LSP decays 
will be rather small. 
Moreover, the leptons and jets from direct stop pair production will be much harder than the lepton 
and jets resulting from $W$ and $Z$ decays in case of LSP pair production. We have checked 
that the hard cuts on leptons and jets in this paper would suppress the background very efficiently. 

\begin{figure}[!htbp]
\begin{center}
{
\epsfxsize=8cm\epsfbox{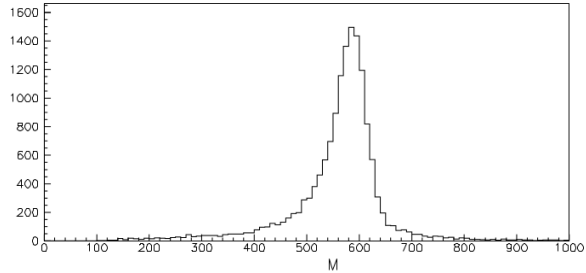}
}
\caption{Invariant mass ($M$) distribution of the muon-jet pair in a dimuon-dijet sample
for $\mlstop = 600$ GeV. See text for details.}
\end{center}
\label{fig:masspeak}
 \end{figure}

One can sharpen the signal further by constructing the invariant mass of the jet-lepton pair. 
In Ref. \cite{adsp2} the reconstruction of the stop mass from the signal events in a particular
channel was illustrated. This is, however, possible even without $b$-tagging.
Suppose we take the two hardest jets in an event and 
combine each of them with the two hard isolated leptons in turn, and keep only those events where
both the invariant masses are close. The result is shown in Fig.\ \ref{fig:masspeak} where 
the difference between two jet-lepton invariant masses is taken to be 20 GeV or less. 
We show the unnormalized distribution of invariant mass of the higher-mass  
muon-jet pair in the dimuon-dijet sample for 
$m_{\stopone} = 600$ GeV. This procedure can be used to further reduce 
the SM backgrounds, if necessary.

\section{Analysis}

In this section we will discuss the APS in the framework of the model 
as defined in Section 2 {\em vis-\`a-vis} the neutrino data.
Apart from the RPV parameters, the neutrino masses and mixing angles depend on several RPC parameters. 
The tree-level contributions, as shown in Eqs.\ (\ref{mtree-c}) and (\ref{mtree-cpr}), 
depend on the combination $C$. As noted earlier, for a bino dominated $\lsp$ and a 
wino dominated $\chonepm$, $C$ is naturally negative, or $C'$ is naturally positive.    
Similarly, the one-loop contributions shown in Eq.\ (\ref{rpv-numass-loop2}) depend on
the SU(2) gauge coupling, the trilinear soft breaking term
$A_b$, $\mu$, $\tan\beta$, and 
the average $b$-squark mass ${m_{\tilde b}}$. The sign of the parameter $K$ defined 
in Eq.\ (\ref{rpv-numass-loop2}) depend on the relative magnitude and sign of $A_b$ and $\mu$.
Note that the second term in Eq.\ (\ref{rpv-numass-loop2}) 
does not involve any further RPC parameters other than those already defined.  

As said earlier, we assume, in the entire analysis, (i) the first two squark 
generations to be much heavier than the third, and (ii) bino-dominated $\lsp$ and 
wino-dominated $\tilde{\chi}^+_1$. 

The numerical inputs are
\be
\mu = 1.762~{\rm TeV}\,,\ \ 
\tan\beta = 10\,,\ \ 
A_t = 3~{\rm TeV}\,,\ \ 
A_b = -5.5~{\rm TeV}\,,\ \ 
m_{\tilde{b}}=2.5~{\rm TeV}\,,\ \ 
M_2-M_1 = 100~{\rm GeV}\,.
\ee
Apart from this, we take $m_{\lsp}-\mlstop = 25$ GeV for the stop LSP case, and 
$\mlstop-m_{\lsp}=25$ and 75 GeV for stop NLSP cases. While $M_1$ is more or less 
fixed by the $\lsp$ mass, $M_2$ is kept close to $M_1$ so that $C'$ in Eq.\ 
(\ref{mtree-cpr}) is large, which, as we will show, is favored by the neutrino
data.

The stop decay width depends on the mixing angle $\theta_t$ defined in 
Eq.\ (\ref{theta-t-def}). This angle (as well as $\mlstop$) is computed from 
the $2 \times 2$ stop mass matrix ${\cal M}^{\tilde t}$.  
We take the diagonal and off-diagonal entries of ${\cal M}^{\tilde t}$ as free parameters. 
The off-diagonal entries are proportional to the stop mixing parameter $X_t=A_t-\mu\cot\beta$; 
this has to be large so that the mass of the lighter Higgs scalar comes out to be in the 
range $126\pm 3$ GeV \cite{uncertainty}. 
This motivates the choice of $A_t$. 
${\cal M}^{\tilde t}_{LL}$, which is the common mass parameter for L-type stop and 
sbottom squarks, is chosen to be large so that the average sbottom mass is 2.5 TeV. 
Thus the relatively small ${\cal M}^{\tilde t}_{RR}$ essentially determines $\mlstop$, 
and this is the parameter that we tune to generate different $\mlstop$. 
Over the entire range of $m_{\stopone}$ corresponding to observable signals, the allowed 
values of $\theta_t$ lie in a narrow range: $0.10 \leq 
\cos\theta_t \leq 0.12$, which is due to the Higgs mass constraint. 
Thus, $\stopone$ is dominantly $\tilde{t}_R$, but even the small $\tilde{t}_L$ 
component makes the RPV channels competitive to the RPC ones for stop NLSP case. 
The mixing angle is of no consequence for stop decays when $\stopone$ is the LSP.

Given the above RPC parameters
the solutions consistent with  the neutrino masses and mixing depend on the interplay of the three 
biliear RPV parameters $\mu_i$ and the three trilinear couplings $\lambda'_{i33}$.
We randomly generate these  parameters varying them over 
a wide range: 
\be
\mu_i \in [10^{-4},5\times 10^{-2}]~{\rm GeV}\,,\ \ 
\lambda'_{i33}\in [2\times 10^{-8},2.5\times 10^{-4}]\,.
\ee
The required parameters are directly read from the SLHA output file 
generated by SUSPECT \cite{suspect}. 
We choose only those points that satisfy all the neutrino constraints and correctly reproduce
the Higgs mass $m_h$ in the allowed region.

The generated mass squared differences and mixing angles must be 
consistent with the neutrino oscillation data \cite{rpv-nu-fit}. At $1\sigma$ level, this 
reads:
\bea
&& \Delta m_{21}^2 =  \left(7.45^{+0.19}_{-0.16}\right)\times 10^{-5}~{\rm eV}^2\,,\ \ 
\Delta m_{31}^2 =  (2.417\pm 0.014)\times 10^{-3}~{\rm eV}^2\,,\nonumber\\
&& \sin^2(\theta_{12}) = 0.306\pm 0.012\,, \ \ 
\sin^2(\theta_{23}) = (0.446\pm 0.008) \oplus 0.593^{+0.027}_{-0.043}\,,\nonumber\\ 
&&\sin^2(\theta_{13}) =0.0231 \pm 0.0019\,.
\label{nu-filter}
\eea
However, the precise determination of $\sin^2\theta_{13}$ has 
severely constrained the parameter space for RPV models\footnote{For 
our discussion, we will assume CP to be conserved in the leptonic sector 
and hence all couplings to be real; however, the strength of the signals 
that we discuss does not depend upon this assumption.}.  
The highly restrictive nature of the neutrino data can be understood from the fact that 
we hardly find any solutions from $10^9$ randomly generated points over the RPV 
parameter space as shown in Eq.\ (\ref{nu-filter}). Therefore, we extend the range to
$3\sigma$ level \cite{rpv-nu-fit} for further analysis \footnote{This does not mean that 
this model is ruled out at $1\sigma$ level.}:
\bea
&& \Delta m_{21}^2 =  [6.98:8.05]\times 10^{-5}~{\rm eV}^2\,,\ \ 
\Delta m_{31}^2 =  [2.247:2.623]\times 10^{-3}~{\rm eV}^2\,,\nonumber\\
&& \sin^2(\theta_{12}) = [0.271:0.346]\,, \ \ 
\sin^2(\theta_{23}) = [0.366:0.663]\,,\ \ 
\sin^2(\theta_{13}) = [0.0173:0.0288]\,.
\label{nu-filter-3sigma}
\eea
The combinations of the RPV parameters which are filtered through the above 
data defines the APS and each member of this set is refered to as a solution.
We also require that the sum of all the neutrino masses must also satisfy
\be
\sum m_{\nu} < 0.22~{\rm eV~at~95\%~CL}\,
\ee
using the data from cosmic microwave background, the baryon acoustic oscillation, and the 
supernova luminosity distance from Hubble space telescope \cite{giusarma}.

The  values of $\lambda'_{i33}$, for every solution, show a hierarchical 
structure; one or two of the couplings will be large compared to the other(s). 
This was also noted in Refs.\ \cite{adspdsp} and \cite{adsp2}. Obviously, 
the number of any particular type of lepton in the final state is directly proportional 
to the square of the RPV coupling. Thus, all possible solutions can be divided into several groups, 
depending upon which lepton(s) are going to be dominant. For example, if $\lambda'_{233}
\gg \lambda'_{133}, \lambda'_{333}$, we will expect only muon signals from such models. 
While there are nine possible leptonic combinations from the two stops, it reduces to 
only three for us ($ee$, $\mu\mu$, and $e\mu$) as we do not simulate final states with $\tau$, 
for reasons discussed before. Let us also note here that the solutions are almost 
equally spread about $\sin^2\theta_{23}=0.5$.

As we will show, for higher stop masses, some solutions are `lost', {\em i.e.} the 
corresponding PBRs in all three dilepton channels ($ee$, $\mu \mu$ and $e \mu$) fall 
below the respective MOPBRs. One might wonder whether $\tau$ detection might help, 
although this possibility cannot be checked as yet for reasons already discussed. 
The pattern of the solutions, however, shows that for every stop mass (both LSP 
and NLSP cases), at least 50\% of the solutions have $\lambda'_{333} > \lambda'_{133}, 
\lambda'_{233}$.  Thus a significant fraction of the so-called `lost' solutions may be 
observed through the $\tau$ channels. For example, with $\mlstop=900$ GeV for the stop 
LSP case and an integrated luminosity of 100 fb$^{-1}$, 144 solutions are lost, out of 
which $\lambda'_{333}$ is largest for 112 solutions. For stop NLSP case, there are 
188 (219) lost solutions with $\mlstop=700$ (900) GeV and $\mlstop-m_{\lsp} = 75$ GeV; 
111 (114) of them have $\lambda'_{333}$ as the largest coupling. Of course, just the 
fact that $\lambda'_{333}$ is largest does not guarantee a detection through the 
$\tau$ channel.

\subsection{${\stopone}$ LSP}

In this subsection we study the APS for models where $\stopone$ is the LSP
and decays entirely through the RPV channels. For our parameter choice, 
$C'$ (Eq.\ (\ref{mtree-cpr})) is positive but $K$ (Eq.\ (\ref{rpv-numass-loop2}))
is negative; for example, 
\begin{equation}
C' = 1.2 \times 10^8~{\rm eV}\,,\ \ 
K = -1.7\times 10^6~{\rm eV}\,,
\end{equation}
for $\mlstop=600$ GeV.
Note that changing the sign of $\mu$ would flip the sign of $K$ but $C'$ is independent of sgn($\mu$).
Later in the paper,
we try to play with these constants \footnote{This is possible as gaugino and 
squark masses are independent parameters, and there can very well be a hierarchy between them.}, 
and it turns out that the above choice is close to the optimal one; no matter how we change the 
coefficients, the number of models that pass the neutrino data filter never has a order-of-magnitude
enhancement. In almost all the cases, it goes down, and sometime goes down rather drastically.

Only the RPV channels are open for $\stopone$ decay, and for each solution the BR
of $\stopone\to \ell_i+b$ is simply proportional to $\left({\lambda'_{i33}}\right)^2$. 
The BRs are independent of $m_{\stopone}$ and $\theta_t$; the dependence 
cancels out {\em only} for the stop LSP case. 

\begin{table}[htbp]
\begin{center}
\begin{tabular}{||c|c|c|c|c|c||}
\hline
$m_{\stopone}$ (GeV) & No.\ of & $N_{ee}$ & $N_{\mu\mu}$ & $N_{e\mu}$ & Lost\\
                     & solutions $N$ &         &                &              & solutions $N_0$\\
\hline
500 & 348 &  73 (104) &   309 (319)  &   239 (260)  &   0 (0) \\
550 & 306 &  48 (59) &   272 (275)  &   176 (209)  &   0 (0) \\
600 & 250 &  41 (43) &   219 (224)  &   105 (136)  &   1 (0) \\
650 & 247 &  40 (41) &   200 (204)  &   92 (132)  &   10 (2) \\
700 & 251 &  46 (50) &   186 (199)  &   64 (111)  &   17 (7) \\
750 & 238 &  31 (32) &   161 (191)  &   35 (73)  &   48 (17) \\
800 & 233 &  17 (25) &   141 (169)  &   6 (41)  &   75 (40) \\
850 & 222 &  14 (19) &   109 (136)  &   0 (11)  &   99 (67) \\
900 & 223 &  8 (20) &     70 (104)  &   1 (1)  &   144 (99) \\
950 & 203 &  0 (6) &      5 (63)  &   0 (0)  &  198 (134) \\
1000 & 199 &  0 (0) &     0 (0)  &   0 (0)  &   199 (199) \\
\hline
\end{tabular}
\end{center}
\caption{Distribution of signals for the various model solutions. Column 2 shows the total number 
of solutions $N$; columns 3-5 show the solutions that result in an observable $ee$, $\mu\mu$, or $e\mu$
signal (PBR  $> P^{\rm min}_{ij}$). The last column shows the number of `lost' solutions $N_0$
for which no signal can be observed. The numbers are for ${\cal L}= 100 (300)$ fb$^{-1}$.}
\label{lsp-stop}
\end{table}

Table \ref{lsp-stop} summarizes our result for this case. The second column gives, 
for different $\mlstop$, the number of solutions $N$
after scanning over $10^9$ randomly generated combinations.  This is only indicative 
of the stringent nature of the neutrino data; $N$ will increase if we employ a finer scan, focussing 
around individual solutions. This number might be contrasted 
with \cite{adsp2}, where a few thousand solutions were obtained for the same number of 
randomly generated points in spite of imposing a tighter constraint of taking the neutrino data at
the $2\sigma$ level. 

It is a bit puzzling to note that $N$, which is apparently independent of
$\mlstop$, decreases significantly for higher values of $\mlstop$. 
This is due to the fact that in this case $M_1$ and $M_2$ must also be 
large to maintain the stop LSP condition. As a result, $C'$ in Eq.\ (\ref{mtree-cpr}) 
goes down while $K$ remains unchanged, thereby reducing the probability of 
satisfying the constraints from neutrino data. 
For higher values of $M_1$ and/or $M_2 - M_1$, 
the number of solutions goes down in general.

Our process is the production of a $\stopone$ pair and its subsequent decay to a pair of leptons 
plus $b$-jets. There are only three leptonic channels ($e e$, $\mu \mu$ and $e \mu$) that we focus upon; for each of these 
channels, the number of solutions $N_{\ell\ell}$ for which the signal is above the MOPBR limit is shown in the 
next three columns of Table \ref{lsp-stop}. The numbers are shown for integrated luminosities 
of 100 and 300 fb$^{-1}$ respectively, the latter within parentheses.  In the last column, we show
the number of `lost' solutions, denoted by $N_0$. 

Note that $N_0$ increases with $m_{\stopone}$; we lose almost half of the solutions for $m_{\stopone}
> 850$ GeV with 100 fb$^{-1}$ and $m_{\stopone} > 900$ GeV with 300 fb$^{-1}$. The higher luminosity 
option is of marginal help, as we have noted before. Above $m_{\stopone} = 950$ GeV, we do not expect 
to see the signals for any solution point. However, $\tau$ detection might be of considerable help.

\subsection{$\stopone$ NLSP}

As discussed in the introduction, for a 
$\stopone$  NLSP with $m_{\stopone} > m_{\lsp}$, there may be various competing RPC 
decay modes of $\stopone$ depending on the parameter space. These modes are: 
(i) $\stopone \to t\lsp$, which is not kinematically allowed if $m_{\stopone}-m_{\lsp} < m_t$, 
(ii) $\stopone \to bW\lsp$, which may be dynamically suppressed if LSP is almost a pure bino, 
(iii) $\stopone\to c\lsp$, which is a loop-induced process,
and (iv) $\stopone \to f\bar{f}' b \lsp$. The  process (ii) swamps the RPV decay channels when
$m_{\stopone} - m_{\lsp} > m_W$ and if the RPV couplings are of the order of $10^{-4}$--$10^{-5}$. 
Thus, we restrict ourselves  to two typical cases: $m_{\stopone} - m_{\lsp} = 25$ and 75 GeV, keeping
$m_{\stopone}$ a free parameter. Thus  only processes (iii) and (iv) may 
possibly be the competing RPC decay channels.  

\begin{table}[!htbp]
\begin{center}
\begin{tabular}{||c|c|c|c|c|c|c|c||}
\hline
$m_{\stopone}$ (GeV) & $m_{\lsp}$ (GeV) & BR (RPV) & No.\ of & $N_{ee}$ & $N_{\mu\mu}$ 
& $N_{e\mu}$ & Missed\\
                     &                  & (\%)     & solutions $N$ &         &                &              & solutions $N_0$\\
\hline
500 & 425 & 6 - 53  & 410 & 53 (59)  & 266 (317) & 49 (82) & 97 (44) \\
    & 475 & 64 - 96 & 384 & 68 (85) & 337 (344) & 230 (248) & 1 (0) \\
    \hline
550 & 475 & 7 - 55 & 388 & 36 (51)  & 174 (251) & 33 (74) & 191 (89) \\
    & 525 & 64 - 97 & 348 & 66 (71)  & 299 (306) & 198 (230)& 2 (2) \\
    \hline
600 & 525 & 7 - 57 & 329 & 16 (33)  & 101 (188) & 15 (39) & 215 (121) \\
    & 575 & 65 - 97 & 299 & 48 (51)  & 249 (260) & 120 (167) & 10 (1) \\
    \hline
650 & 575 & 8 - 63 & 269 & 17 (23)  & 52 (77) & 2 (18) & 200 (173) \\
    & 625 & 61 - 97 & 250 & 36 (41)   & 207 (219) & 65 (99) & 12 (1) \\
    \hline
700 & 625 & 6 - 69 & 256 & 19 (21)  & 50 (67) & 0 (4) & 188 (168)\\
    & 675 & 63 - 97 & 247 & 37 (41)   & 184 (203) & 59 (99) & 28 (7) \\
    \hline
750 & 675 & 8 - 68 & 238 & 13 (14)  & 39 (51) & 0 (5) & 187 (173)\\
    & 725 & 63 - 98 &  251 & 39 (47)  & 161 (185) & 28 (72) & 52 (19) \\
    \hline
800 & 725 & 8 - 72 & 241 & 8 (16)  & 38 (58) & 0 (0) & 195 (168)\\
    & 775 & 65 - 98 &  225 &  20 (24) & 133 (158) & 5 (37) & 71 (45) \\
    \hline
850 & 775 & 6 - 72 & 238 & 2 (7)  & 16 (43)& 0 (0)& 223 (188)\\
    & 825 & 66 - 98 &  233 & 13 (17)  & 102 (133) & 0 (3)& 119 (83) \\
    \hline
900 & 825 & 8 - 74 & 219 & 0 (2)  & 0 (21)& 0 (0)& 219 (196) \\
    & 875 & 69 - 98 &  222 & 6 (11)  & 59 (93) & 0 (0) & 157 (118) \\
    \hline
950 & 875 & 8 - 74 & 223 & 0 (0)  & 0 (0)& 0 (0)& 223 (223) \\
    & 925 & 64 - 99 &  223 & 0 (6)  & 2 (58) & 0 (0) & 221 (159) \\
    \hline
1000& 925 & 8 - 81 & 212 & 0 (0)  & 0 (0)& 0 (0)& 212 (212) \\
    & 975 & 67 - 98 &  203 & 0 (0)  & 0 (0) & 0 (0) & 203 (203)\\
\hline
\end{tabular}
\end{center}
\caption{Same as Table \ref{lsp-stop} but for the stop NLSP case. 
For each stop mass, two cases are shown where the LSP $\lsp$ is 25 (75) GeV below $m_{\stopone}$. 
The branching fractions for both stops decaying through RPV channels are also shown. 
The numbers are for ${\cal L}= 100 (300)$ fb$^{-1}$.}
\label{nlsp-stop}
\end{table}

The RPC parameters are fixed as before. For a given stop mass, the stop mixing 
angle $\theta_t$ is no longer a constant like the LSP case, because of the presence of RPC channels. 
However, $\cos\theta_t$ still remains small, almost at about the same range as in the LSP case. 

 Our results are shown in Table \ref{nlsp-stop}, which is 
similar to Table \ref{lsp-stop}, except that the second column shows the 
LSP mass and the third column shows the BR of the stop through RPV channels. 
The remaining columns 4--8 carry the 
same information as colums 2--6 of Table \ref{lsp-stop}.

Several points are to be noted here. First, $N$ is fairly insensitive to  
$m_{\stopone} - m_{\lsp}$ for fixed $\mlstop$. However, $N$ decreases with increasing  
$\mlstop$ for reasons discussed in the last subsection. Second, the probability 
of RPV decays (by which we mean both the $\stopone$s will decay to $\ell+b$) goes 
down with increasing $m_{\stopone} - m_{\lsp}$, and this is true for all 
values of $\mlstop$. This is because the RPC channels start to open up with 
increasing phase space. 
Third, the `reach' for the stop NLSP case\footnote{`Reach' is used here in a bit cavalier
way. What we mean is the stop mass above which we lose more than half of the solutions.}
is in general a bit lower compared to the stop LSP case, 
which is again due to the dilution of the RPV branching fractions from the RPC channels.

\subsection{Parameter variation}

In this subsection we vary some of the key RPC parameters individually, keeping others fixed.
The rationale is to check whether our choice is the optimal or close to optimal one, or whether 
there exists some other benchmark with allows a significantly larger number of solutions. Of course the 
RPV parameters $\mu_i$ and $\lambda'_{i33}$ are generated randomly and scanned over their entire
range. We show the entire variation for $\mlstop = 600$ GeV. For the stop NLSP case, we display 
results only for  
$m_{\stopone}-m_{\lsp}=25$ GeV. 
In all tables in this subsection, the first numerical columns represent our benchmark scenario, followed
by the variations.

\subsubsection{Variation of $\mu$}
Table \ref{tab:mu-var} shows the effect of changing $\mu$ by $\pm 500$ GeV from our benchmark 
of 1762 GeV. While the number 
of solutions increase (decrease) for lower (higher) values of $\mu$, the change is not so
drastic.
One might wonder why there should be a variation at all, as the tree-level neutrino 
mass matrix depends on the ratios $\mu_i/\mu$ and $\mu_j/\mu$. However, $\mu$ also appears 
in  the first term of Eq.\ (\ref{rpv-numass-loop2}) in a nontrivial way, and the interplay 
determines the number of solutions.

\begin{table}[htbp]
\begin{center}
\begin{tabular}{|c|c|c|c|}
\hline
$\mu$ (GeV) & 1762 & 1262 & 2262\\
\hline
No.\ of  & 299 & 618 & 260 \\
solutions & (250) & (614) & (248) \\
\hline
\end{tabular}
\end{center}
\caption{Variation of $\mu$. We show the number of solutions in $10^9$ tries that pass the neutrino filter. 
The numbers in parentheses are for the stop LSP case.}
\label{tab:mu-var}
\end{table}

\subsubsection{Variation of $M_2-M_1$}
The results are shown in Table \ref{tab:m2m1-var}. Note that there is only a modest change of N as we vary 
our benchmark mass difference of 100 GeV. However, low values of $M_2-M_1$, which increase $C'$, 
help towards more solutions.

\begin{table}[htbp]
\begin{center}
\begin{tabular}{|c|c|c|c|}
\hline
$(M_2 - M_1)$ (GeV) & 100 & 300  &500 \\
\hline
No.\ of    & 299 & 232 & 219 \\
solutions & (250) & (239)& (221)\\
\hline
\end{tabular}
\end{center}
\caption{Variation of $M_2-M_1$.
The numbers in parentheses are for the stop LSP case.}
\label{tab:m2m1-var}
\end{table}

\subsubsection{Variation of average bottom squark mass}

Entries in $M^{\lambda'\lambda'}$ are inversely proportional to $m^2_{\tilde {b}}$, 
and those in $M^{\mu\lambda'}$ are inversely proportional to $m_{\tilde{b}}$, the 
average $b$-squark mass.
The number of solutions decrease when $m_{\tilde b}$ goes down. When 
$m_{\tilde b}$ is increased, the behavior is 
different for $\stopone$ LSP and NLSP cases, but the number of solutions remain 
at the same order of magnitude. Thus, our choice of $m_{\tilde b}$ is optimal for stop 
NLSP and close to optimal for stop LSP case. 
For the NLSP case, smaller $m_{\tilde{b}}$ enforces smaller $\lambda'_{i33}$ to generate the 
same solutions, so the RPV BRs also go down. 

\begin{table}[htbp]
\begin{center}
\begin{tabular}{|c|c|c|c|}
\hline
$m_{\tilde b}$ (GeV) & 2500 & 1500 &3000\\
\hline
No.\ of & 299 & 70 & 200 \\
solutions & (250) & (109) & (499)\\
\hline
\end{tabular}
\end{center}
\caption{Variation of $m_{\tilde b}$. The numbers in parentheses are for the stop LSP case.}
\label{tab:mbr-var}
\end{table}

\subsubsection{Variation of $C'$ and $K$}

Next, we vary the  parameters $C'$ in Eq.\ (\ref{mtree-cpr}) and $K$ in Eq. (\ref{rpv-numass-loop2}),
one at a time. 
We see from Table \ref{tab:c-var} that our choice of $C'$ is the optimal one; both the 
allowed number of solutions and the RPV branching ratios fall as we move by one order of magnitude 
to either side. Note that the variation of $C'$ affects only the tree-level mass term, 
unlike the variation of $\mu$, which affects all terms. 
Also, the relative values of $C'$ and $K$ are important to generate correct atmospheric as well
as solar splittings. Identical conclusions appear from the variation of $K$, 
see Table \ref{tab:k-var}, where we have shown our results also by flipping the sign of 
$\mu$. 
Even though we have not considered the variation of the $\mu_i\lambda'_{i33}$ term, this exercise alone 
shows that the neutrino mass hierarchy is strongly preferred to be a normal one.

\begin{table}[htbp]
\begin{center}
\begin{tabular}{|c|c|c|c|}
\hline
$C'$ (eV)& $1.2\times 10^8$ & $1.2\times 10^7$ & $1.2\times 10^9$\\
\hline
No.\ of & 299 & 115 & 8 \\
solutions& (250) & (107) & (11) \\
\hline
\end{tabular}
\end{center}
\caption{Variation of the tree-level coefficient $C'$ of the neutrino mass matrix.
The numbers in parentheses are for the stop LSP case.}
\label{tab:c-var}
\end{table}

\begin{table}[htbp]
\begin{center}
\begin{tabular}{|c|c|c|c|c|}
\hline
$K$ (eV) & $-1.7\times 10^{6} $ & $-1.7\times 10^5$ & $-1.7\times 10^7$ & $8.9\times 10^5$ \\
\hline
No.\ of  & 299 & 114 & 9 & 99\\
solutions& (250) & (2) & (10) & (83)\\
\hline
\end{tabular}
\end{center}
\caption{Variation of the loop-level coefficient $K$ of the neutrino mass matrix. 
The numbers in parentheses are for the stop LSP case. The 
last column is when the sign of $\mu$ is flipped: $\mu = -1762$ GeV.}
\label{tab:k-var}
\end{table}

Before we end, let us mention that the
rotation from $\{H_2,L_i\}$ basis to $\{H'_2,L'_i\}$ basis can in principle induce 
charged lepton-chargino, and neutrino-neutralino mixing. They can 
affect the decays of the NLSP, $\stopone$, or the LSP, $\lsp$. 
For example the RPC vertex $\stopone$--$b$--chargino may 
lead to additional lepton number violating couplings of the $\stopone$ due 
to chargino-lepton mixings. Similar induced couplings may arise from 
$\stopone$--$t$--neutralino coupling due to mixing 
in the neutralino-neutrino sector (these couplings may be relevant only 
if the $\stopone$ is significantly heavier than the top quark). The mixing 
factors which would suppress the induced couplings can be estimated from 
the 5 $\times$ 5 (7 $\times$ 7) chargino-lepton (neutralino-neutrino) 
mass matrix. The estimated value is ${\cal O}(\mu_i / \mu) \sim 10^{-7}$. Since 
the largest $\mu_i$ allowed by the oscillation data is ${\cal O}(10^{-4})$ 
GeV, the mixing factors are estimated to be ${\cal O}(10^{-7})$ or smaller. 
Moreover, the induced couplings will be additionally suppressed by gauge 
or Yukawa couplings. On the other hand the smallest $\lambda'_{i33}$ coupling 
contributing to $\stopone$ decay consistent with the nuetrino oscillation data is 
${\cal O}(10^{-5})$. Thus the $\stopone$-NLSP BRs computed by considering 
$\lambda'_{i33}$ driven decays only are quite reliable. 

We now justify the above estimates by some numerical results, analogous to those 
obtained in \cite{adsp1}. We numerically diagonalize  the 
mass matrices in the chargino-lepton or the neutralino-neutrino sector for
all combinations  of RPV parameters allowed by the oscillation data. We 
find that the chargino-charged lepton mixing is always extremely tiny;
A typical amplitude for 
finding a charge lepton mass eigenstate in a chargino is
$1.6 \times 10^{-6}$. The corresponding $\lambda'_{133}$, responsible for 
the $\stopone$ or LSP decay signal, is $1.2 \times 10^{-4}$. Similarly
the maximum amplitude for
finding a neutrino  mass eigenstate in any neutralino is
$1.5 \times 10^{-6}$. The corresponding $\lambda'_{133}$ is $1.2 \times 
10^{-4}$. The above results are for $m_{\lspone} = 500 $ GeV but this trend is seen 
over the entire range  of parameters considered in this paper. 


\section{Summary}

In conclusion, we reiterate that the current neutrino data severely constrain the 
RPV generated neutrino mass models. In this paper, we have discussed a particular type 
of such models, where the neutrino mass matrix is generated by three bilinear and 
three trilinear $\lambda^{\prime}_{i33}$-type couplings. The severity of the constraint 
can be gauged from the fact that out of a billion randomly generated sets of the above 
six parameters, only a few hundred at the most are found to be consistent with the data. 
While the actual number of solutions depends on the RPC parameters, we have explicitly 
checked that our choice of those parameters is close to optimal. We have intentionally 
refrained from making an absolute optimal choice of the RPC parameters, just to keep our 
analysis on the conservative side. We have also kept our analysis on the conservative side, by 
considering only electrons and muons in the final state, and not employing $b$-tagging,
as $\tau$- and $b$-tagging efficiencies at LHC-14 TeV are not yet publicly available. 
Even with such a conservative analysis, we get a lot of interesting conclusions.

The model considered in this paper may be tested through
the RPV decays of the lighter stop squark, $\stopone$, provided it is the 
LSP or the NLSP. These scenarios with a relatively light $\stopone$ are now well 
motivated since the mixing parameter $X_t$ in the stop sector is likely to be large 
in view of the measured value of $m_h \approx$ 126 GeV, resulting in a rather 
light stop mass eigenstate. For a stop pair produced at LHC-14 TeV, 
we consider their RPV decays into the channels  $\ell_i\ell_j + 2$ jets. 
Using {\tt PYTHIA} based simulations, we obtain  model independent 
estimates of the minimum observable product branching ratios as a function of $\mlstop$. 
In contrast to an earlier work by two of us, we show that a favourable signal may be obtained 
without $b$-tagging. Following this procedure one can also reconstruct the mass of 
$\stopone$ from the lepton-jet invariant mass distributions, which in turn reveals 
the lepton number violating nature of the underlying stop decays.

We then consider the RPV models consistent with neutrino data and compute the product 
branching ratios for each of them. We find that for lower values of $\mlstop$, signals 
from almost all the models are observable; on the other hand, the fraction of observable 
models goes down with increasing $\mlstop$, and for 
$\mlstop$ around 1 TeV, no signal can be seen for any of the allowed models. We 
have checked that many of the lost solutions correspond to $\lambda^{\prime}_{333} >
\lambda^{\prime}_{133}, \, \lambda^{\prime}_{233}$, especially for moderately heavy 
$\stopone$. Once the techniques for handling $\tau$ rich final states at LHC-14 TeV 
experiments and the corresponding background estimates are available, the inclusion 
of these states will improve the observability of neutrino mass models.

\section{Acknowledgements} 

R.B. acknowledges Council for Scientific 
and Industrial Research, Govt.\ of India, and  
A.K. acknowledges Department of Science and Technology, Govt.\ of India, 
and Council for Scientific 
and Industrial Research, Govt.\ of India for research support. 
A.D acknowledges the award of an emeritus Senior Scientist Position by the Indian
National Science Academy.


\end{document}